\documentclass{article}
\usepackage[T1]{fontenc}
\usepackage{a4wide}
\usepackage{graphics}

\makeatletter

\providecommand{\LyX}{L\kern-.1667em\lower.25em\hbox{Y}\kern-.125emX\@}

\newcommand{\lyxrightaddress}[1]{
  \par {\raggedleft \begin{tabular}{l}\ignorespaces
  #1
  \end{tabular}
  \vspace{1.4em}
  \par}
}

\usepackage[T1]{fontenc}

\makeatletter

\usepackage[T1]{fontenc}

\makeatletter

\usepackage[T1]{fontenc}

\makeatletter

\usepackage[T1]{fontenc}

\makeatletter

\usepackage[T1]{fontenc}


\makeatother
\makeatother
\makeatother
\makeatother

\begin{document}

\title{Charged Particle Fluctuation \\
as Signal of the Dynamics in Heavy Ion Processes}

\author{Fritz W. Bopp and Johannes Ranft \\
   \\
   Universität Siegen, Fachbereich Physik, \\
 D--57068 Siegen, Germany }

\maketitle
\vspace*{-8cm}

\lyxrightaddress{SI-01-3 (hep-ph/0105192, revised version)}

\vspace*{+7cm}

\begin{abstract}
We compare the dispersion of the charges in a central rapidity box according
to the Dual Parton Model with the predictions of statistical models. Significant
deviations are found in heavy ion collisions at RHIC and LHC energies. Hence
the charged particle fluctuations should provide a clear signal of the dynamics
of heavy ion processes. They should allow to directly measure the degree of
thermalization in a quantitative way.
\end{abstract}

\section{Introduction}

In the analysis of the hadronic multi-particle production (for a recent review
see \cite{jacob99}) a key observation has been the local compensation of charge~\cite{idschok73, whitmore76,foa75,brandelik81,misra83}
. In fixed target hadron-hadron experiments all charges in the forward region
could be determined. It was therefore possible to consider the dispersion of
the charge fluctuation between a forward and backward hemisphere~\cite{chouyang73,sivers74,bialas75,phua77,lam77,chiu76,diasdedeus77,jezabek77}
. In this way significant results could be obtained already in the 70'ties at
comparatively low energies. The charge fluctuations connected to the soft hadronic
part of the reactions were found to involve only a restricted rapidity range.
This observation limited the applicability of statistical models to rather local
fluctuations (see e.g. \cite{ranft75}). 

Good agreement was obtained in cluster model calculations (the clusters were
used to parameterize the effect of resonances). Early versions assumed neutral
clusters and obtained reasonable results. A predicted Quigg-Thomas relation
for the forward-backward charge fluctuation across the rapidity boundary \( y \)
\begin{equation}
\label{eq-1}
<\delta Q_{>y}^{2}>=<(Q_{>y}-<Q_{>y}>)^{2}>=c\cdot dN_{\mathrm{non}\, \mathrm{leading}\, \mathrm{charge}}/dy
\end{equation}
 was satisfied (for a review see e.g. \cite{ranft75}). The agreement could
be improved if charged clusters with mesonic Regge-exchanges~\cite{bopp78}
were allowed. As the charge structure is quite similar the same good agreement
can be expected for the Dual Parton Model. Using the Dual Parton Model code
DPMJET~\cite{DPMJET} we explicitely checked that this is indeed the case. For
\( pp \)-scattering at laboratory energies of 205 GeV good agreement with the
data~\cite{kafka75} was obtained. The Quigg-Thomas relation is satisfied in
the calculation with \( c=0.70 \) which compares with an experimentally preferred
value of \( c=0.72 \).

In heavy ion scattering charge flow measurements should be analogously decisive.
It is a central question of an unbiased analysis whether the charges are distributed
just randomly or whether there is some of the dynamics left influencing the
flow of quantum numbers. This is not an impractical conjecture. In heavy ion
experiments the charge distribution of the particle contained in a central box
with a given rapidity range \( [-y_{\mathrm{max}.}+y_{\mathrm{max}.}] \) can
be measured and the dispersion of this distribution
\begin{equation}
\label{eq-2}
<\delta Q^{2}>=<(Q-<Q>)^{2}>
\end{equation}
can be obtained to sufficient accuracy even if some of the charges are misidentified.
For sufficiently large gaps this quantity contains information about long range
charge flow. In comparison to the dispersion of the forward (resp. backward)
charge which have been studied at FNAL energies, the charge distribution in
a central box (having two borders) can be expected to require twice the rapidity
range.

It was proposed to use the quantity (1) to distinguish between particles emerging
from an equilibrized quark-gluon gas or from an equilibrized hadron gas~\cite{heinz00,jeon00,jeon99}.
In a hadron gas each particle species in the box is taken essentially poissonian.
In a central region at high energies where the relative size of the box is small
and where the average charge flow can be ignored, one obtains a simple relation
for particles like pions with charges \( 0 \) and \( \pm 1 \) 

\begin{equation}
\label{eq-3}
<\delta Q^{2}>=<N_{\mathrm{charged}}>.
\end{equation}
 The inclusion of resonances reduces hadron gas prediction by a significant
factor taken ~\cite{jeon00,bleicher00} to be around \( 0.7 \). It is now argued
in the cited papers that this relation would change in a quark gluon gas to
\begin{equation}
\label{eq-4}
<\delta Q^{2}>=\sum _{i}q_{i}^{2}<N_{i}>=0.19<N_{\mathrm{charged}}>
\end{equation}
where \( q_{i} \) are the charges of the various quark species and where again
a central region is considered. The coefficient on the right was calculated~\cite{jeon00}
for a two flavor plasma in a thermodynamical consideration which predicts various
quark and gluon contributions with suitable assumptions. A largely empirical
final charged multiplicity \( N_{\mathrm{charged}}=\frac{2}{3}(N_{\mathrm{glue}}+1.2N_{\mathrm{quark}}+1.2N_{\mathrm{antiquark}}) \)
was used. 

There is a number of sources of systematic errors in the above comparison between
the QGP and the hadron gas. The result strongly depends on what one takes as
primordial and what as secondary particles. Considering these uncertainties
we follow the conclusion of Fia{\l }kowski's papers ~\cite{fialkowski00} that
a clear cut distinction between the hadron- and the quark gluon gas is rather
unlikely. This does not eliminate the interest in the dispersion. 

In the next section we discuss various possible measures to observe such fluctuations.
In section 3 a simple interpretation of the dispersion in terms of quark lines
is outlined. An obtained proportionality suggests to compare the dispersion
to the particle density instead of the enclosed total particle multiplicity.
This comparison is presented in section 4 in the framework of a Dual Parton
model Monte Carlo code (DPMJET). Modeling the statistical charge distribution
by randomizing charges, the charge transfer dispersion is shown to allow a clear
distinction between string models and equilibrium approaches starting with RHIC
energies. These predictions for RHIC and LHC collisions are presented in section
5.

\section{Various Measures for Charge Fluctuations }

\vspace{0.3cm}
For the analysis of the charge structure several quantities were discussed in
the recent literature. It was proposed to look at the particles within a suitable
box of size \( \Delta y \) and to measure just the mean standard deviation
of the ratio \( R \) of positive to negative particles

\begin{equation}
\label{eq-5}
<\delta R^{2}>=<\left( \frac{N_{+}}{N_{-}}\: -<\frac{N_{+}}{N_{-}}>\right) ^{2}>
\end{equation}
or the quantity \( F \) 
\begin{equation}
\label{eq-6}
<\delta F^{2}>=<\left( \frac{Q}{N_{\mathrm{charged}}}-<\frac{Q}{N_{\mathrm{charged}}}>\right) ^{2}>
\end{equation}
 where \( Q=N_{+}-N_{-} \) is the charge in the box. 

The motivation for choosing these ratios was to reduce the dependence of multiplicity
fluctuations caused p.e. by variations in the impact parameter. In the region
of interest (experimental measurements of large nuclei at high energies with
a suitable centrality trigger) the charge component of the fluctuations strongly
dominates and envisioned cancellation in density fluctuations is not important. 

In this region the charge fluctuation and the proposed quantities (i.e. (3),(5)
and (6)) are equivalent. They are simply connected by the following relations~\cite{jeon00}
\begin{equation}
\label{eq-7}
<N_{\mathrm{charged}}><\delta R^{2}>=4\: <N_{\mathrm{charged}}><\delta F^{2}>=4\cdot \frac{<\delta Q^{2}>}{<N_{\mathrm{charged}}>}.
\end{equation}

To examine the new quantities and the range where these relations hold, all
three quantities were calculated in the Dual Parton model implementation DPMJET~\cite{DPMJET}
. For the most central \( 5\% \) Pb-Pb scattering at LHC energies (\( \sqrt{s}=6000 \)
A GeV) there is indeed a perfect agreement between all three quantities as shown
in figure~1. This agreement stays true for analogous Pb-Pb data at RHIC energies
(\( \sqrt{s}=200 \) A GeV) .\footnote{
The extreme region above \( \Delta y>10 \) is not relevant as it is not accessible
to foreseeable experiments. Obviously, in the limit where the box extends over
the total kinematic range, where \( n_{+}-n_{-\: }\rightarrow \: Z_{1}+Z_{2} \)
is fixed the dispersions \( <\delta R^{2}> \)and \( <\delta F^{2}> \)is dominated
by pure multiplicity fluctuations.
} 

Outside the region of interest of central heavy ion collisions the proposed
alternatives have problems. They are not suitable for smaller \( \Delta y \)
boxes in less dense events, as they are actually undefined (\( 0/0 \) or \( \infty  \))
if no suitable particles in the corresponding box exist.
\begin{figure*}
{\par\centering \resizebox*{0.48\columnwidth}{0.4\textheight}{\includegraphics{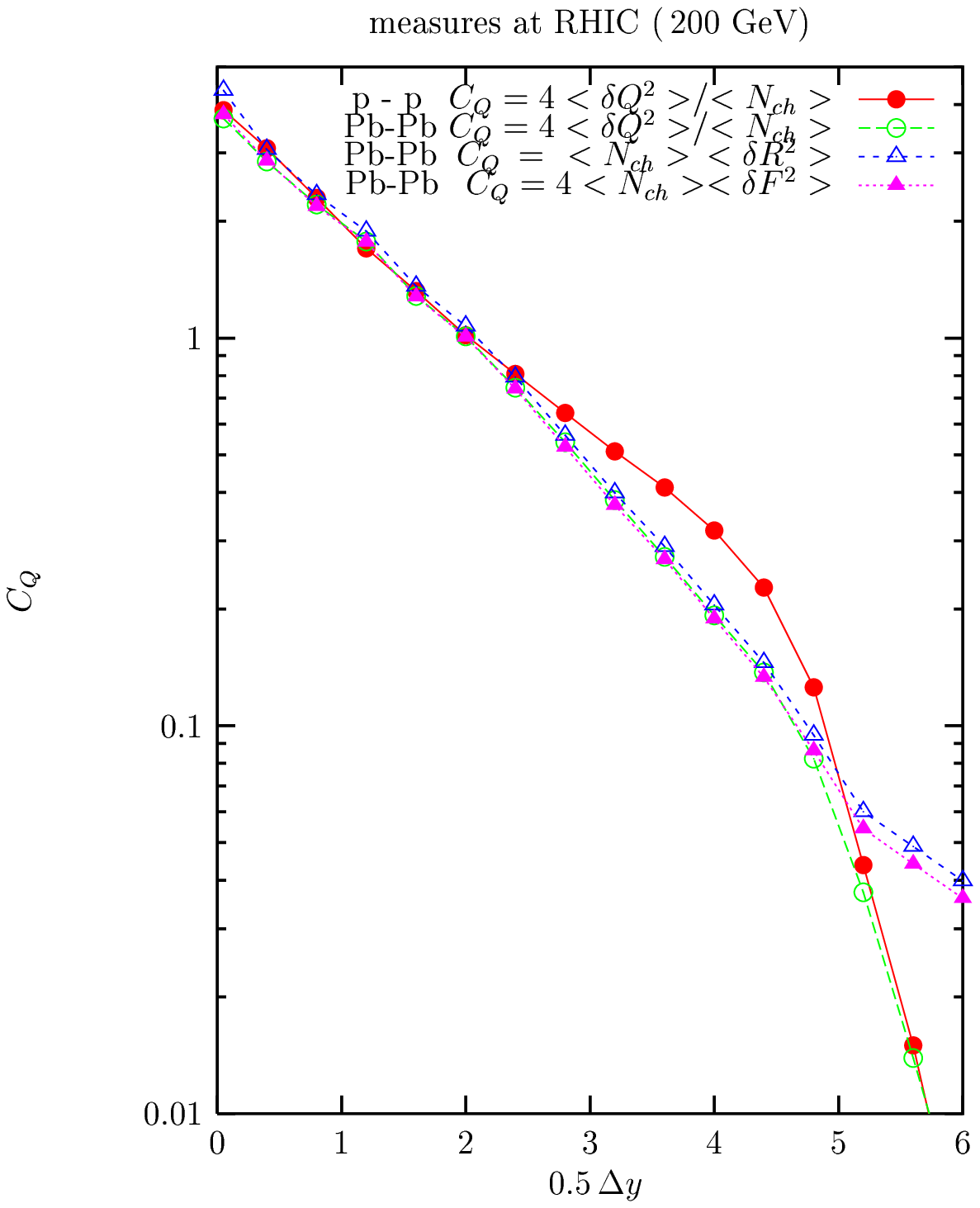}} \resizebox*{0.48\columnwidth}{0.4\textheight}{\includegraphics{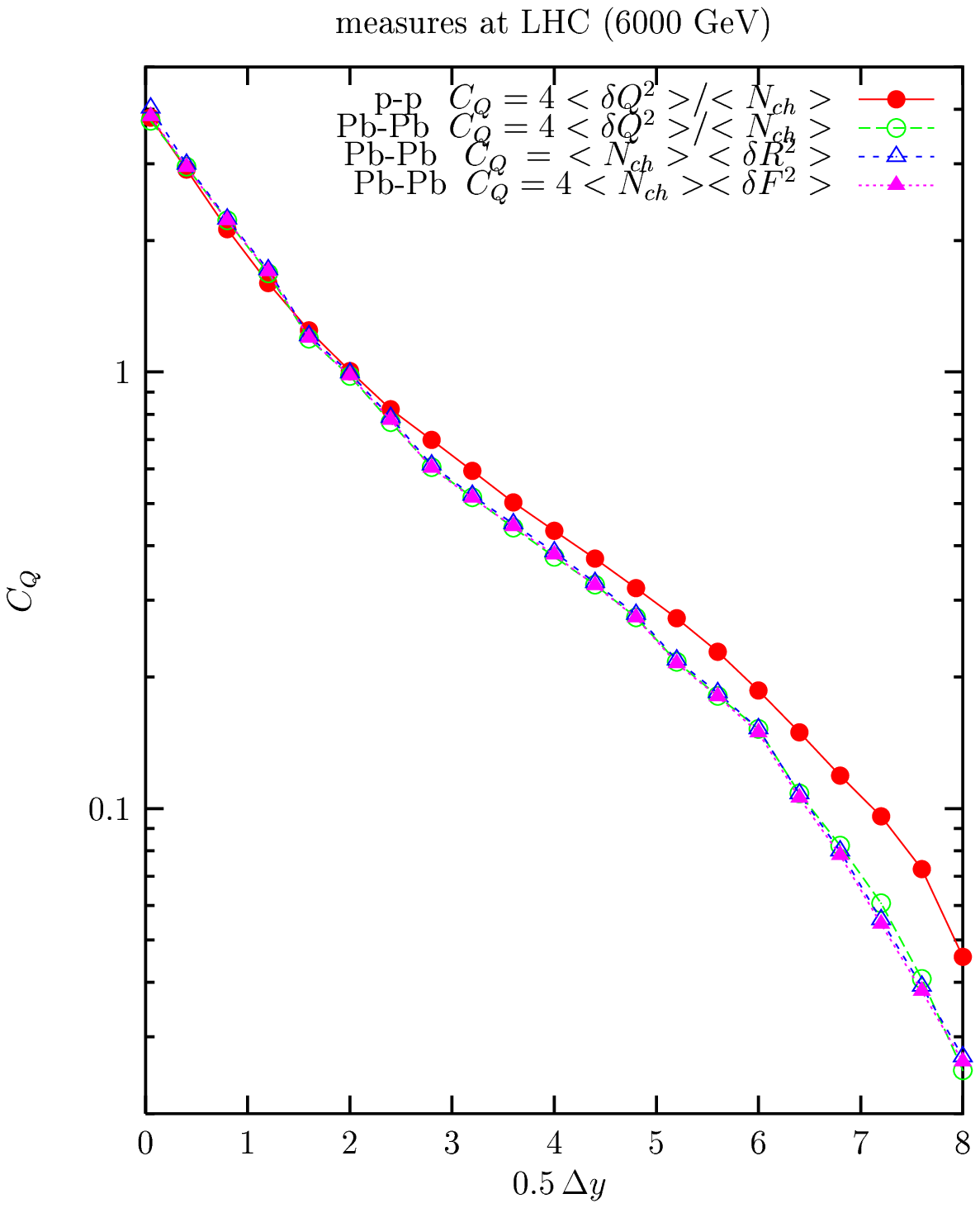}} \par}

\caption{Charge fluctuations for the most central \protect\( 5\%\protect \) Pb-Pb scattering
at RHIC energies (\protect\( \sqrt{s}=200\protect \) A GeV and at LHC energies
(\protect\( \sqrt{s}=6000\protect \) A GeV). Also shown are corresponding data
for p-p scattering}
\end{figure*}

We tried to fix the problem by ignoring undefined contributions but the result
was not satisfactory as their mutual relation (equation 7) is lost with the
appearance of such terms. Specifically for the most central \( 5\% \) S-S scattering
at RHIC energies, the agreement is no longer good and for the minimum bias S-S
scattering or for proton-proton scattering at these energies the agreement is
lost. 

From a string model point of view any conclusion will strongly depend on a comparison
of central processes with minimum bias and proton-proton events. Our explicite
Monte Carlo calculation indicates that for this purpose the more regular behaved~\cite{initial}
dispersion of the net charge distribution \( <\delta Q^{2}> \) might be best
suited.

The slightly flatter distribution for the proton proton scattering is easy to
understand. One has to realize that a glauber model does not involve just as
a simple superposition. In heavy ion scattering each incoming nucleon will participate
in several scattering processes. In consequence there will be more and somewhat
shorter strings.

For none of the variables significant differences between rapidity and pseudo-rapidity
boxes were observed. We did not investigate the influence of particles with
small transverse momentum which may escape detection in present experiments.

\section{A Simple Relation between the Quark Line Structure and Fluctuations in the
Charge Flow}

To visualize the meaning of charge flow measurements it is helpful to introduce
a general factorization hypothesis. It is not exact but it usually holds to
good accuracy. It postulates that the light flavor structure of an arbitrary
hadronic amplitude can be described simply by an overall factor, in which the
contribution from individual quark lines factorize \footnote{
The hypothesis is based on the exchange degeneracy of octet and singlet Regge
trajectories effectively changing the \( SU(N_{\mathrm{effective}}) \) flavor
symmetry to an \( U(N_{\mathrm{effective}}) \) symmetry in which this relation
is exactly valid. Corrections to the hypothesis originate in the special behavior
of the masses of the lowest lying mesons of the trajectories, which is especially
significant in the pseudo-scalar sector i.e. between the \( \pi _{0} \) and
the \( \eta  \) meson. This introduces an anti-correlation between flavors
on neighboring quarks which can be ignored in considerations concerning on long
range charge transfers.

If a higher accuracy is desired the hypothesis can be restricted to primary
particles which are less sensitive to these masses\cite{aurenche77,bopp78}.
The ``secondary'' charges (produced in pairs during the decay of large primordial
particles) have then to be considered extra using a Quigg-Thomas relation\cite{quigg73,quigg75,baier74}
\( <\delta Q^{2}(y)>=\sigma \frac{1}{2}\rho _{\mathrm{charged}\: \mathrm{secondary}}(y) \)
where \( \sigma \approx 1 \) . 
}. 

The hypothesis can be used to obtain the following generalization of the Quigg-Thomas
relation \cite{baier74,aurenche77,bopp78,bopp81}. It states that the correlation
of the charges \( Q(y_{1}) \) and \( Q(y_{2}) \), which are exchanged during
the scattering process across two kinematic boundaries, is just 
\begin{equation}
\label{eq-8}
<\left\{ Q(y_{1})-<Q(y_{1})>\right\} \left\{ Q(y_{2})-<Q(y_{2})>\right\} >=n_{\mathrm{common}\: \mathrm{lines}}<(q-<q>)^{2}>.
\end{equation}
where \( n_{\mathrm{common}\: \mathrm{lines}} \) counts the number of quark
lines intersecting both borders and \( q \) is the charge of the quark on such
a line. Depending on the flavor distribution average values \( <(q-<q>)^{2}=0.22\cdots 0.25 \)
are obtained. 

Most observables of charge fluctuations can be expressed with this basic correlation.
Our fluctuation of the charges within a \( [-y_{\mathrm{max}.},+y_{\mathrm{max}.}] \)
box contains a combination of three such correlations. Using (\ref{eq-8}) for
each contribution the dispersion of the charges in a box subtracts to 
\begin{equation}
\label{eq-9}
<\delta Q[\mathrm{box}]^{2}>=n_{\mathrm{lines}\: \mathrm{entering}\: \mathrm{box}}<(q-<q>)^{2}
\end{equation}
where \( n_{\mathrm{lines}\: \mathrm{entering}\: \mathrm{box}} \) is the number
of quark lines entering the box.

\section{Calculation of the Dispersion of the Charge Distribution within a Box}

Let us consider the prediction of a thermodynamic model in more detail. In the
thermodynamic limit with an infinite reservoir outside and a finite number of
quarks inside, all quark lines will connect to the outside as shown in figure~2.
\begin{figure}
{\par\centering \resizebox*{!}{0.15\textheight}{\includegraphics{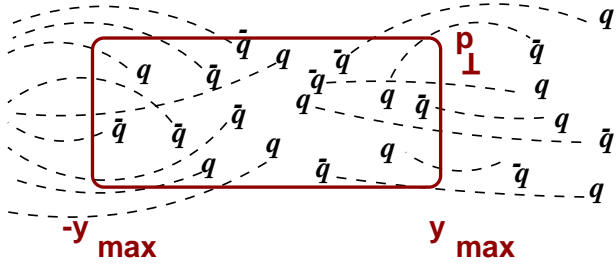}} \par}

\caption{Quark lines entering the box in the thermodynamic limit }
\end{figure}
The dispersion of the charge transfer is therefore proportional to the total
number of particles inside. In the ``hadron gas'' all particles contain two
independent quarks each contributing with roughly \( 1/4 \) yielding the estimate
of equation~2. For the ``quark gluon gas'' only one quark or gluon of each
hadron originates in a non-local process. The other partons needed for the hadronization
are assumed to be short range so that for a box of a certain size their contribution
can possibly be ignored. In this way the charge transfer is drastically reduced.
Obviously there are several refinements to this simple picture.

Let us consider the limit of a tiny box. Looking only at the first order in
\( \Delta y \) one trivially obtains

\begin{equation}
\label{eq-10}
<\delta Q^{2}>/<N_{\mathrm{charged}}>=1
\end{equation}
which corresponds to the hadron gas value. 

If the box size increases to one or two units of rapidity on each side this
ratio will typically decrease, as most models contain a short range component
in the charge fluctuations. One particular short range fluctuation might be
caused by the hadronization of partons of the quark gas discussed above. The
decreasing is however not very distinctive. Common to many models are secondary
interactions which involve decay processes and comover interaction. In hadron
hadron scattering processes such short range correlations are known to play
a significant role and there is no reason not to expect such correlations for
the heavy ion case.

After a box size passed the short range the decisive region starts. In all global
statistical models \cite{heinz00,jeon99,jeon00} the ratio will have to reach
now a flat value. Only if the box involves a significant part of the total rapidity,
charge conservation will force the ratio to drop by a correction factor

\begin{equation}
\label{eq-11}
\mathrm{factor}=\left( \int _{y_{\mathrm{max}.}}^{Y_{\mathrm{kin}.\mathrm{max}.}}\rho _{\mathrm{charge}}^{\mathrm{new}}dy\right) /\left( \int _{0}^{Y_{\mathrm{kin}.\mathrm{max}.}}\rho _{\mathrm{charge}}^{\mathrm{new}}dy\right) \propto 1-y_{\mathrm{max}.}/Y_{\mathrm{kin}.\mathrm{max}.}
\end{equation}
.

This is different in string models. The model calculations (figure~1) with its
rapid fall off indicate a manifestly different behavior. It is a direct consequence
of the local compensation of charge contained in string models. The effect is
illustrated in figure~3 in which only quark lines are shown which intersect
the boundary and contribute to the charge flow.
\begin{figure}
{\par\centering \resizebox*{!}{0.15\textheight}{\includegraphics{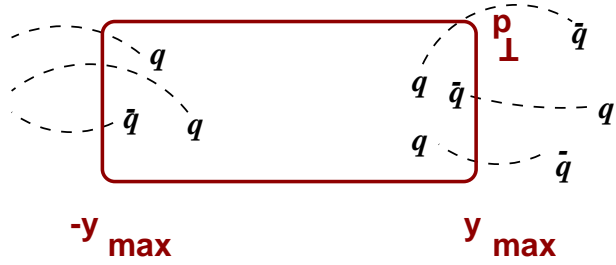}} \par}

\caption{Quark lines entering the box with local compensation of charge}
\end{figure}
 Now local compensation of charge allows only a contribution of lines originating
around the boundaries. If the distance is larger than the range of charge compensation
the dispersion will no longer increase with the box size. The total contribution
will now be just proportional to the density of the particles at the boundaries
\begin{equation}
\label{eq-12}
<\delta Q^{2}>\, \propto \rho _{\mathrm{charged}}(y_{\mathrm{max}.}).
\end{equation}
 It just counts the number of strings.

This resulting scaling, which is indeed very similar to the relation (1), which
was already found in the 70'ties,  is illustrated in a comparison between both
quantities in (12) shown in figure~4 for RHIC and LHC energies.
\begin{figure}
{\par\centering \resizebox*{!}{0.4\textheight}{\includegraphics{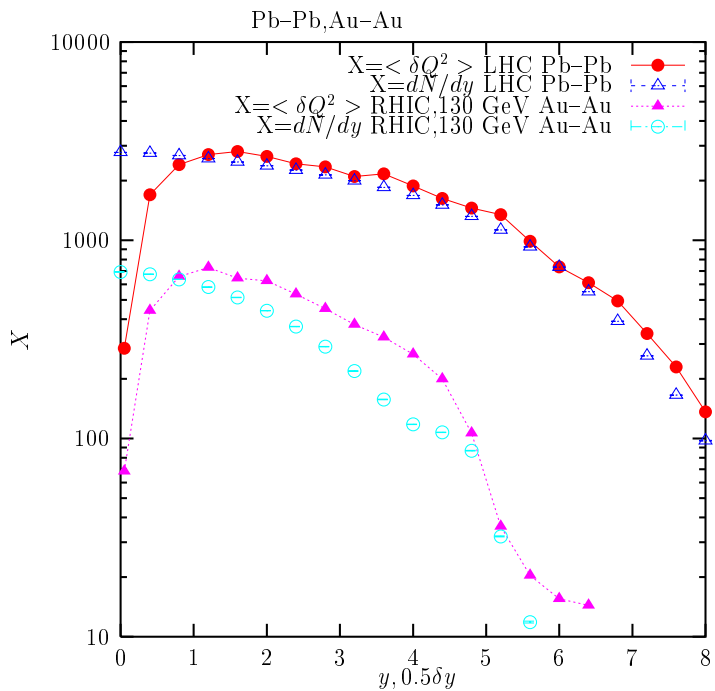}} \par}

\caption{Comparison of the dispersion of the charge distribution with the density on
the boundary of the considered box for central gold gold resp. lead lead scattering
at RHIC and LHC energies.}
\end{figure}
 The agreement is comparable to the proton-proton case shown in figure~5.
\begin{figure}
{\par\centering \resizebox*{!}{0.4\textheight}{\includegraphics{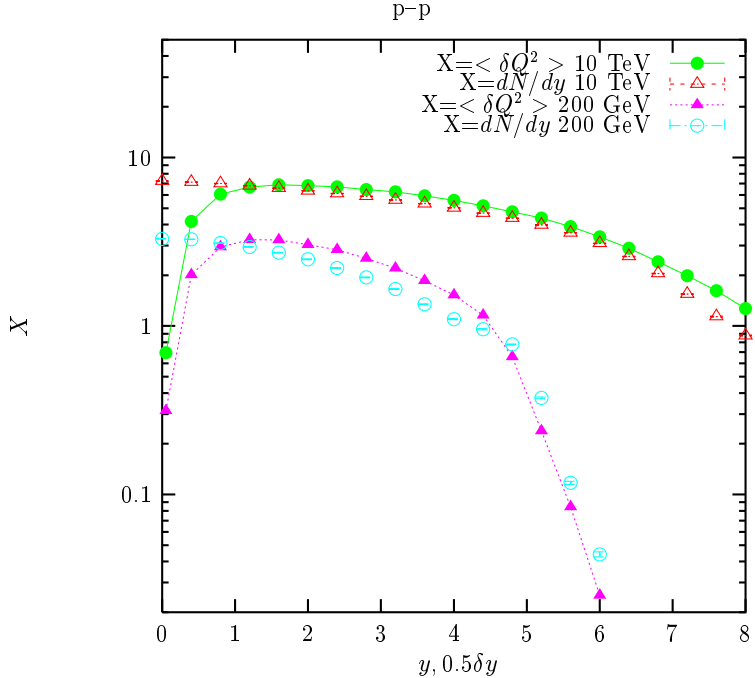}} \par}

\caption{Comparison of the dispersion of the charge distribution with the density on
the boundary of the considered box for proton-proton scattering at RHIC and
LHC energies.}
\end{figure}
 The proportionality is expected to hold for a gap with roughly \( \frac{1}{2}\delta y>1 \)
. For smaller boxes some of the quark lines seen in the density do not contribute
as they intersect both boundaries. For large rapidity sizes there is a minor
increase from the leading charge flow \( Q_{L} \) originating in the incoming
particles. In a more careful consideration\cite{bopp78} one can subtract this
contribution
\begin{equation}
\label{eq-13}
<\delta Q^{2}>_{\mathrm{leading}\: \mathrm{charge}\: \mathrm{migration}}=<Q_{L}>(1-<Q_{L}>)
\end{equation}
 and concentrate truly on central fluctuations. 
\vspace{0.3cm}

The prediction for the proportionality factor for the case of mere short range
fluctuations would be roughly a factor one (see footnote~1). In string models
primordial particles are responsible for a longer range charge transfer coming
from the contributions of the quark resp. diquark fragmentation chains. Taking
everything together one obtains
\begin{equation}
\label{eq-14}
<\delta Q^{2}>=\sum _{\mathrm{left}+\mathrm{right}}\{\: n_{\mathrm{strings}}\cdot 2<(q-<q>)^{2}>+\: \sigma \: \frac{1}{2}\: \rho _{\mathrm{charged}\, \mathrm{secondary}}(y)\: \}
\end{equation}
 where \( n_{\mathrm{strings}}=\rho _{\mathrm{charged}\, \mathrm{primary}}/\rho _{\mathrm{single}\, \mathrm{string}} \)
is the number of strings. The width of the local fluctuations \( \sigma  \)
is roughly unity.

\section*{Expectations for RHIC and LHC collisions}

In a recent publication Bleicher, Jeon, Koch~\cite{bleicher00} pointed out
that the overall charge conservation cannot be ignored at SPS energies. They
showed that their string model prediction\footnote{
In the energy range above \( \sqrt{s}=5 \) GeV the UrQMD code used by them
is described~\cite{bleicher99} to be dominated by string fragmentation. 
} coincides with the expectation of a statistical model of hadrons and that the
considered measure is therefore not sufficiently decisive in the considered
energy range. Our string model DPMJET supports this conclusion for the SPS energy
range as it obtains the same central-box charge fluctuations. While forward-backward
hemisphere charge fluctuations were meaningful in the FNAL-SPS energy region,
the fluctuations of charges into a central box contain two borders and require
a correspondingly doubled rapidity range. 

It was argued~\cite{bleicher00} that the experimental results should be \char`\"{}purified\char`\"{}
to account for charge conservation. In our opinion a sufficiently reliable estimate
of this factor is not available and and the implementation of the charge conservation
has to stay on the model side. The estimate of Bleicher, Jeon, Koch is based
on equation~11 . For \( <\delta R^{2}> \) and \( <\delta F^{2}> \) the corresponding
relation holds only to first order, which seems at least on the formal side
not sufficient. Even for \( <\delta Q^{2}> \) it should be taken with care.
One can easily underestimate the effect of charge conservation, as even in statistical
models \cite{ranft72} not all charged particles might be fully mixed in. The
leading particle often exhibit a special behaviour.

To obtain an estimate in a reference model with statistical fluctuation we a
posteriori randomized charges in final states obtained with DPMJET. A similar
procedure to create a reference sample could be directly applied to experimental
data. To conserve energy and momentum absolutely accurately it was done in our
calculation separately for pions, kaons and nucleons. The result is shown in
figure~6 for RHIC and LHC energies for proton-proton and central lead-lead collisions.
To check consistency we employed the proposed correction factor 
\[
1-\int _{0}^{y_{\mathrm{max}.}}\rho _{\mathrm{charge}}^{}\: dy/\int _{0}^{Y_{\mathrm{kin}.\mathrm{max}.}}\rho _{\mathrm{charge}}^{}\: dy\]
and indeed obtained the flat distribution with the expected ``hadron gas''
value. 

\begin{figure}
{\par\centering \resizebox*{0.48\columnwidth}{0.45\textheight}{\includegraphics{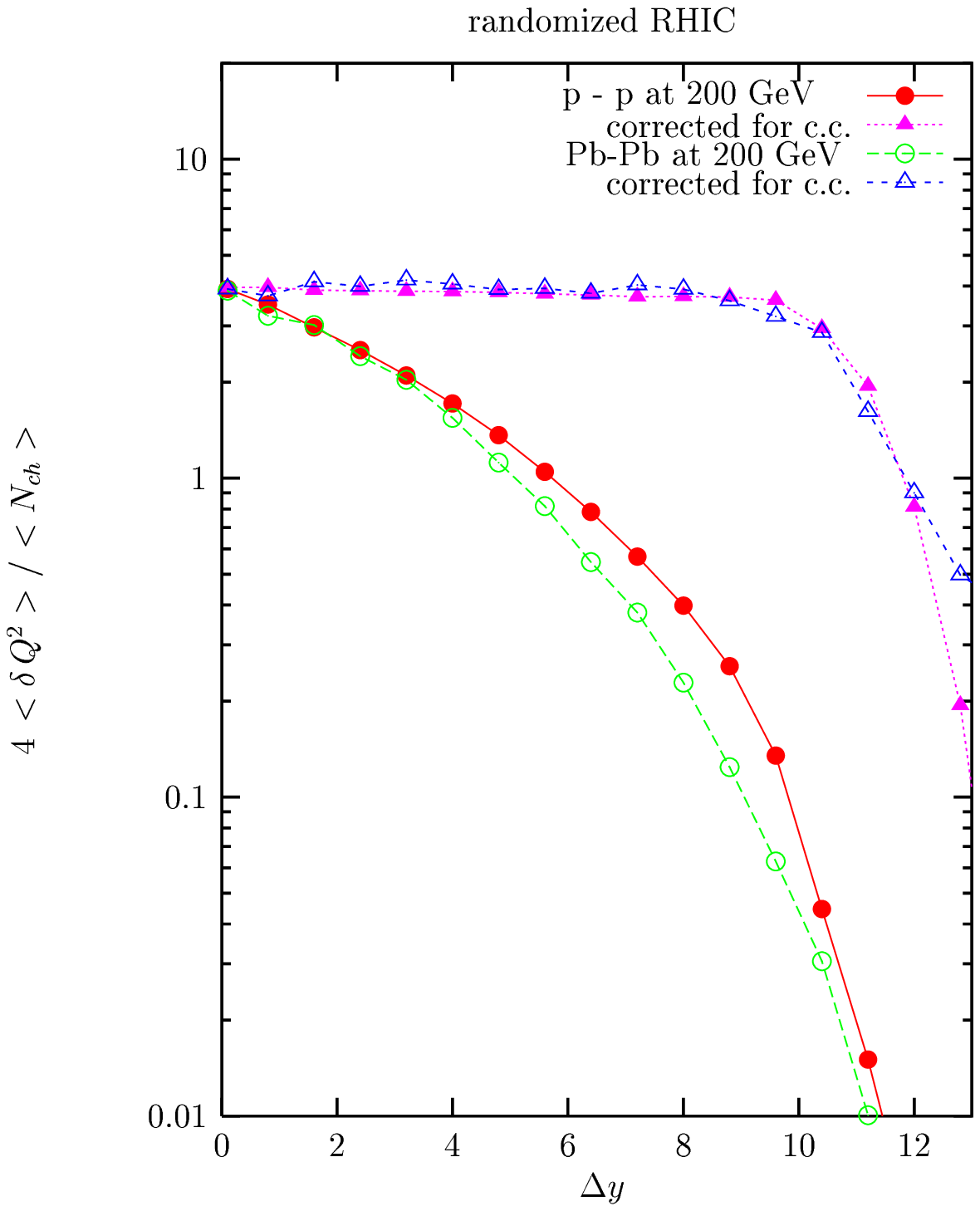}} 
\resizebox*{0.48\columnwidth}{0.45\textheight}{\includegraphics{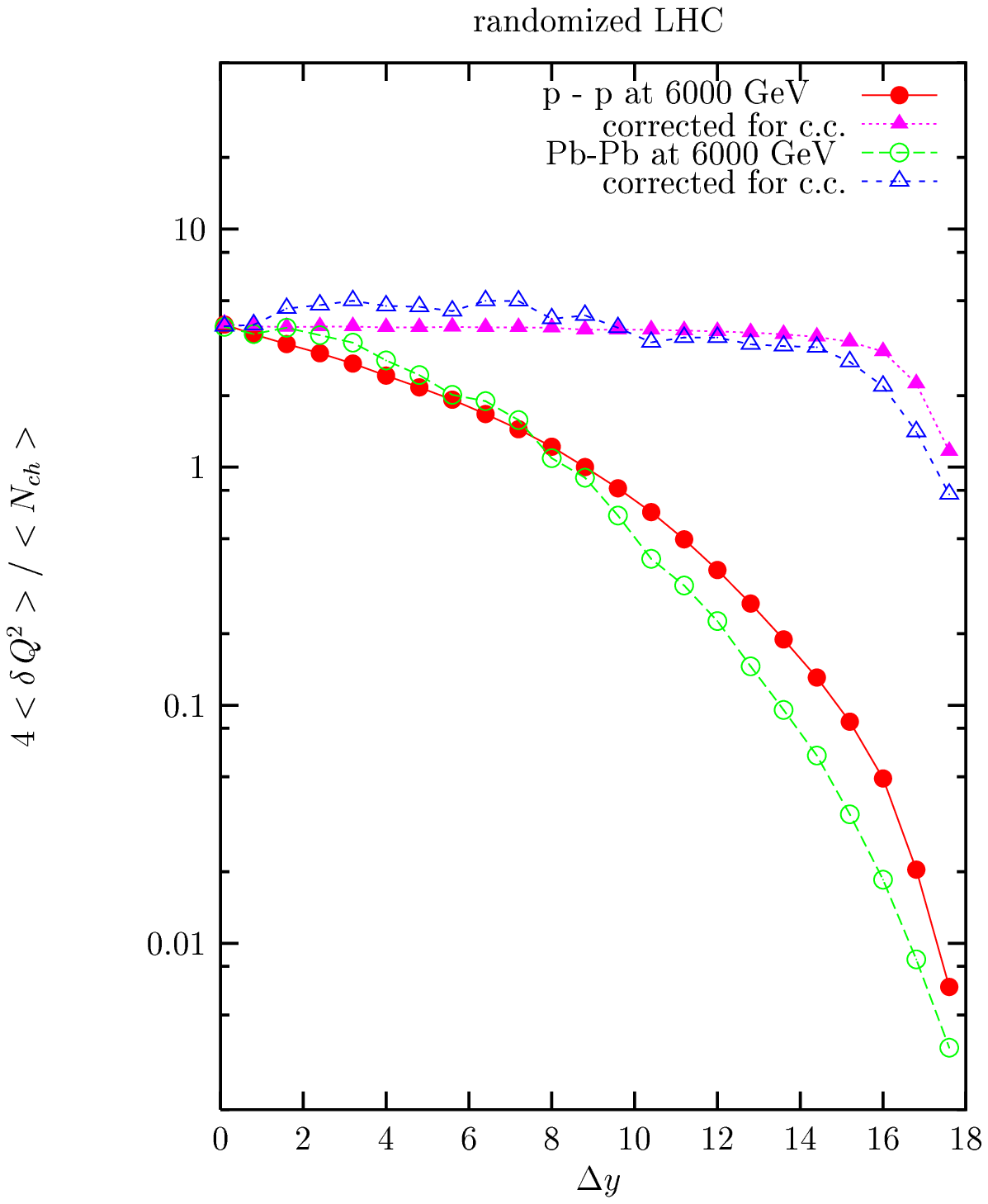}} \par}

\caption{Charge fluctuations with a posteriori randomized charges for p-p scattering
and the most central \protect\( 5\%\protect \) in Pb-Pb scattering at RHIC
energies (\protect\( \sqrt{s}=200\protect \) A GeV) and at LHC energies (\protect\( \sqrt{s}=6000\protect \)
A GeV). The results are also shown with a correction factor to account for the
overall charge conservation.}
\end{figure}

Taking the DPMJET string model and the randomized ``hadron gas'' version as
extreme cases we can investigate the decisive power of the measure. As shown
in figure~7 we find that there is a measurable distinction at RHIC energies
and sizable at LHC. 
\begin{figure}
{\par\centering \resizebox*{0.48\columnwidth}{0.4\textheight}{\includegraphics{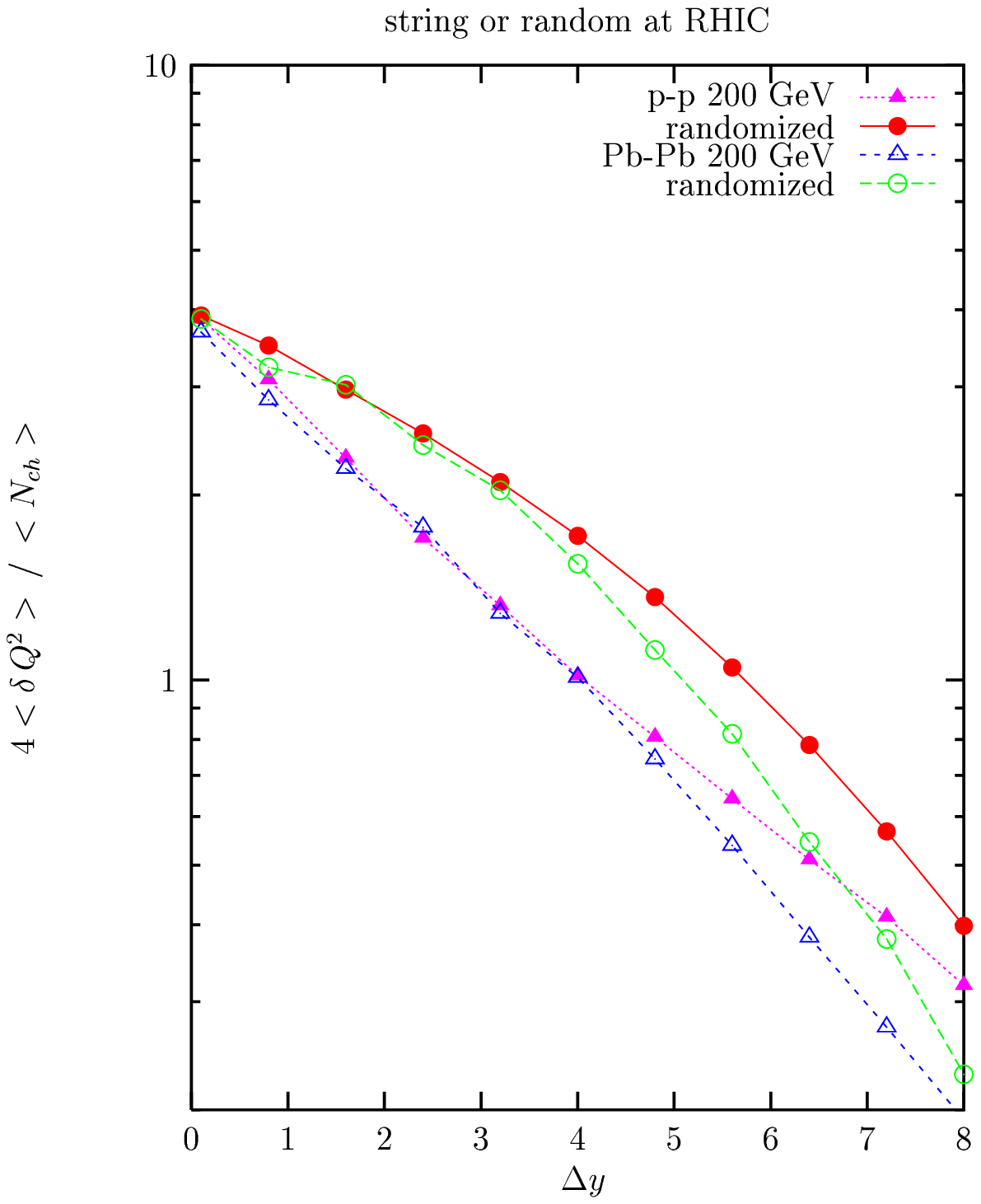}} 
\resizebox*{0.48\columnwidth}{0.4\textheight}{\includegraphics{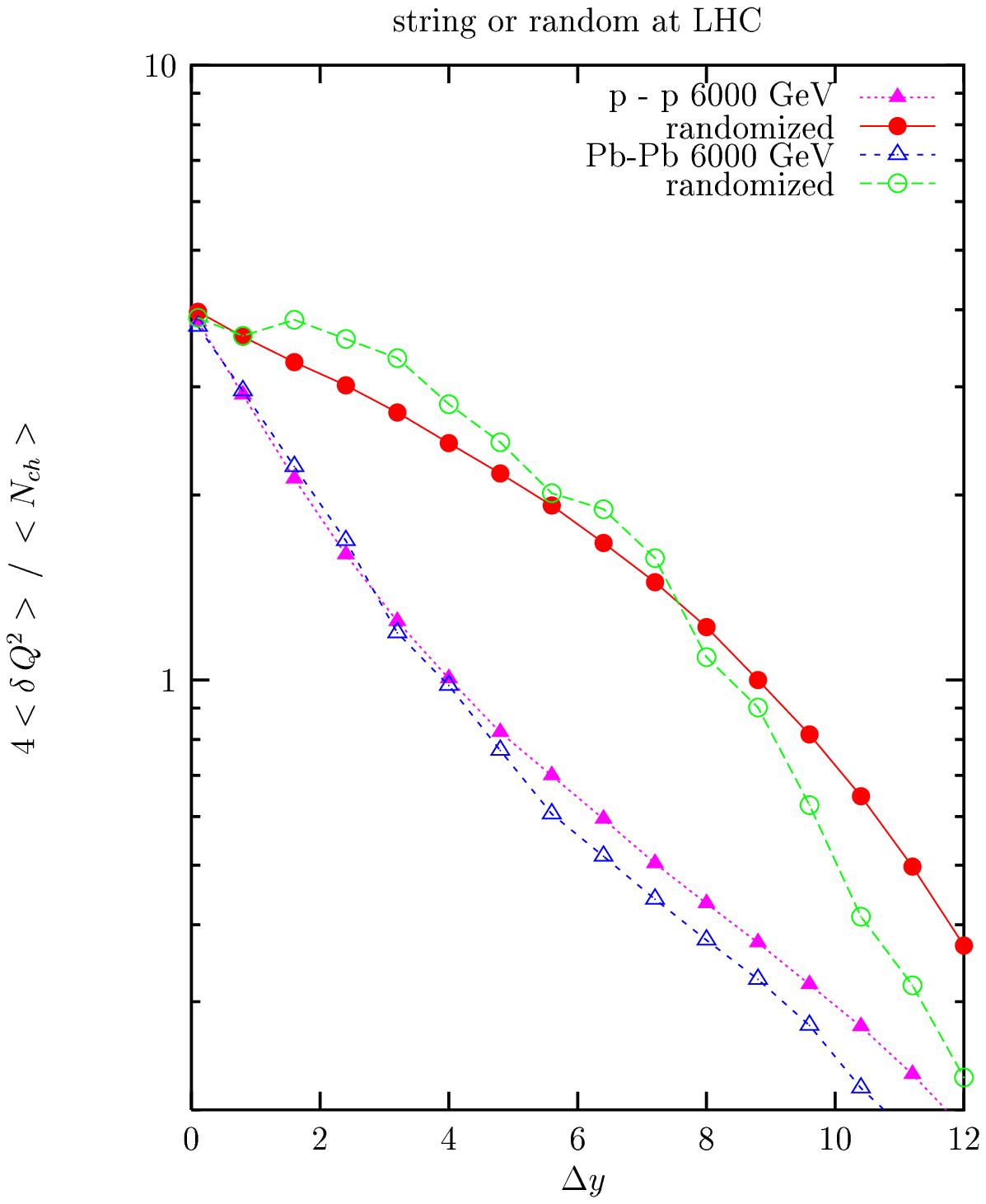}} \par}

\caption{Comparison of the charge fluctuations obtained in a string model DPMJET with
a model using a posteriori randomized charges for p-p scattering and the most
central \protect\( 5\%\protect \) in Pb-Pb scattering at RHIC energies (\protect\( \sqrt{s}=200\protect \)
A GeV) and at LHC energies (\protect\( \sqrt{s}=6000\protect \) A GeV). }
\end{figure}

The similarity of p-p and Pb-Pb scattering is not surprising. The distinction
between both cases is expected from the difference in collective effects. The
data for p-p scattering are known to follow the string models, while interaction
of comovers, or medium range or complete equilibration will move the curve upward
to a more statistical situation. These effects are presently outside of the
model. A measured charge correlation between both extremes will directly reflect
the underlying new physics. 
\begin{figure}
{\par\centering \resizebox*{0.6\columnwidth}{0.4\textheight}{\includegraphics{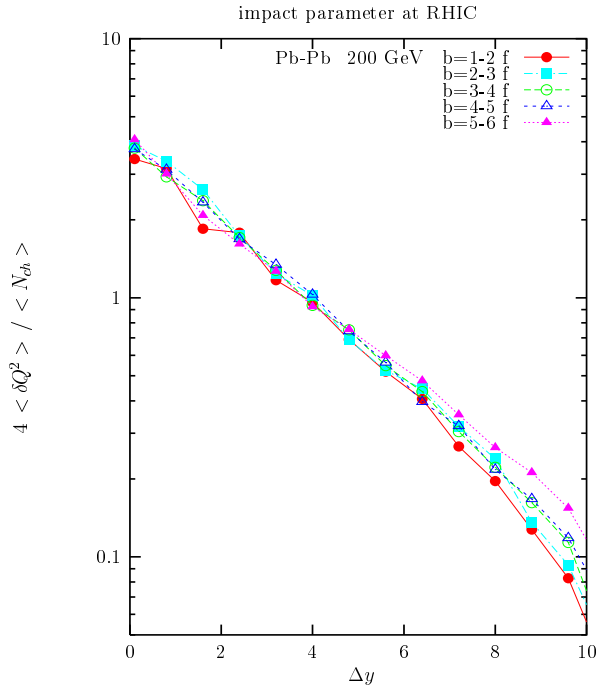}} \par}

\caption{The b dependence of the charge fluctuations obtained in a string model DPMJET
for Pb-Pb scattering at RHIC energies (\protect\( \sqrt{s}=200\protect \) A
GeV). }
\end{figure}

A similar result is obtained when the dependence on the centrality is studied.
Without collective effects no such dependence is expected and found in the model
calculation as can be seen in figure~8 (\( b \) is the impact parameter). It
should be stressed that this experimentally measurable centrality dependence
allows to directly observe collective effects without reference to model calculations
and underlying concepts.

\section*{Conclusion}

In the paper we demonstrated that the dispersion of the charge distribution
in a central box of varying size is an extremely powerful measure. Within the
string model calculation the dispersion seen in relation to the spectra shows
no significant difference between simple proton-proton scattering and central
lead lead scattering even though both quantities change roughly by a factor
of 400. 

The dispersion allows to clearly distinguish between conventional string models
and hadronic thermal models for a rapidity range available at RHIC energies.
In many models the truth is expected to lie somewhere in between. It is a quite
reasonable hope that the situation can be positioned in a quantitative way.

\section*{Acknowledgments}

F.W. Bopp acknowledges partial support from the INTAS grant 97-31696.

\end{document}